\documentclass[sn-mathphys-num]{sn-jnl}% Math and Physical Sciences Numbered Reference Style 

\usepackage{graphicx}%
\usepackage{multirow}%
\usepackage{amsmath,amssymb,amsfonts}%
\usepackage{amsthm}%
\usepackage{mathrsfs}%
\usepackage[title]{appendix}%
\usepackage{xcolor}%
\usepackage{textcomp}%
\usepackage{manyfoot}%
\usepackage{booktabs}%
\usepackage{algorithm}%
\usepackage{algorithmicx}%
\usepackage{algpseudocode}%
\usepackage{listings}%
\usepackage{verbatim}

\begin{document}

\title[Article Title]{Medusa 84 SiH - A novel high Selectivity Electron Beam Resist for Diamond Quantum Technologies}

\author[1]{\fnm{Oliver Roman} \sur{Opaluch}}\email{o.opaluch@rptu.de}

\author[1]{\fnm{Sebastian} \sur{Westrich}}

\author[1]{\fnm{Nimba} \sur{Oshnik}}

\author[2]{\fnm{Philipp} \sur{Fuchs}}

\author[2]{\fnm{Jan} \sur{Fait}}

\author[3,4]{\fnm{Sandra} \sur{Wolff}}

\author[5]{\fnm{Harry} \sur{Biller}}

\author[5]{\fnm{Mandy} \sur{Sendel}}

\author[2]{\fnm{Christoph} \sur{Becher}}

\author*[1]{\fnm{Elke} \sur{Neu}}\email{nruffing@rptu.de}

\affil[1]{\orgdiv{Department of Physics and Research Center OPTIMAS}, \orgname{RPTU University Kaiserslautern Landau}, \orgaddress{\street{Erwin-Schrödinger-Str. 46}, \city{Kaiserslautern}, \postcode{67663}, \country{Germany}}}

\affil[2]{\orgdiv{Fachrichtung Physik}, \orgname{Universität des Saarlandes}, \orgaddress{\street{Campus E2.6}, \city{Saarbrücken}, \postcode{66123}, \country{Germany}}}

\affil[3]{\orgdiv{Department of Physics}, \orgname{RPTU University Kaiserslautern Landau}, \orgaddress{\street{Erwin-Schrödinger-Str. 46}, \city{Kaiserslautern}, \postcode{67663}, \country{Germany}}}

\affil[4]{\orgdiv{Nano Structuring Center}, \orgname{RPTU University Kaiserslautern Landau}, \orgaddress{\street{Erwin-Schrödinger-Straße}, \city{Kaiserslautern}, \postcode{67663}, \country{Germany}}}

\affil[5]{\orgname{Allresist GmbH}, \orgaddress{\street{Am Biotop 14}, \city{Strausberg}, \postcode{15344}, \country{Germany}}}

\abstract{We investigate the novel electron beam resist Medusa 84 SiH by Allresist GmbH (Germany) for nanostructuring of single crystal diamond and its effects on the spin properties of nitrogen vacancy (NV) centers in nanopillars as prototypes for photonic structures. We find contrast curves comparable to previously employed resists (Hydrogensilsequioxane FOx). We present a selectivity for diamond etching of 11 to 12. Using an adhesion-promoting silicon interlayer potentially enables fabrication yields of up to 96\%. We measure $T_2$ times of up to $\sim$25 $\mu$s before and after processing, demonstrating that the manufactured structures are usable for diamond-based quantum sensing.}

\keywords{diamond, lithography (deposition), reactive ion etching, scanning electron microscopy (SEM), electron beam resist}

%%\pacs[MSC Classification]{35A01, 65L10, 65L12, 65L20, 65L70}

\maketitle

\section{Introduction}
\label{Introduction}

In recent decades, diamond has become a versatile material platform for many applications, including electronics, photonics, and, especially, quantum technologies \cite{atature2018material,shandilya2022diamond,orphal2024coherent}. In particular, diamond has several outstanding characteristics that make it suitable for quantum technologies: within its wide bandgap, it hosts atom-like energy states of optically active point defects called color centers. Often, these color centers have long-lived, highly coherent, and optically readable spin states and are thus highly suitable for spin-based quantum sensors. In this context, first it is crucial to embed color centers in diamond nanostructures that enhance the fluorescence collection and, in turn, the optical spin state readout \cite{hausmann2010fabrication,babinec2010diamond,appel2016fabrication,neu2014photonic,Qnami2021}. Second, tip-shaped nanostructures allow nanoscale imaging with single color centers by scanning the diamond tip in proximity to the sample under investigation.\\
In addition, diamond is chemically inert, making it a suitable sensor material for harsh chemical environments. However, considering nanofabrication, chemical inertness poses a challenge; obtaining nanostructures typically requires plasma etching of diamond in a high-density plasma in an inductively coupled plasma reactive ion etcher (ICP-RIE). These plasmas contain oxygen and use a high bias to obtain steep sidewalls of diamond nanostructures. Such plasma conditions, however, pose high demands on the mask material used for etching; strong mask erosion might occur and consequently limit the etch depth and aspect ratio of diamond nanostructures. While metal masks fabricated in lift-off processes can withstand the plasma etching condition, they might lead to non-optimal side walls of the nanostructures. For many previous works, a mask which is obtained directly via electron beam lithography has been used: Hydrogen Silsesquioxane (HSQ)-based electron beam resists are able to withstand harsh conditions, as the chemical composition of the exposed and developed resist closely resembles SiO$_2$ ("spin-on-glass").\\
HSQ excels as a versatile resist due to its low line edge roughness, its ability to planarize structured surfaces, and its low dielectric constant. Applications of HSQ range from molecular precursor material for the manufacturing of silicon-based nanomaterials \cite{hessel2006hydrogen,barry2011synthesis,milliken2021tailoring}, over dielectric interlayers and protective coatings up to resists for many established lithography processes. It shows significant etch resistance to O$_2$ plasma, suitable for various material systems \cite{mollard2002hsq}. HSQ's performance during lithography and as etch mask are beneficial for research, allowing its use for structuring novel substrate materials directly or by defining hard mask geometries. With applications on silicon in CMOS fabrication \cite{van2000hydrogen, trellenkamp2003patterning}, over GaAs / InGaAsP in photonics \cite{lauvernier2005realization,dylewicz2010fabrication}, silicon-on-insulator (SOI) in sensors fabrication \cite{gnan2008fabrication} up to more exotic materials like indium tin oxide (ITO) for lasers \cite{solard2020optimal} or Ge$_2$Sb$_2$Te$_5$ as phase change non-volatile memory \cite{nam2007electron}, HSQ proved its versatility and relevancy for nowadays research. In case of diamond, however, a moderate selectivity is obtained and consequently resist layers of several hundred nanometers are used.\\
A main product used in literature \cite{hausmann2010fabrication,babinec2010diamond} as well as in our previous work \cite{appel2016fabrication,radtke2019plasma,radtke2019reliable,neu2014photonic} that satisfies these specifications is Flowable Oxide (FOx, DuPont, formerly DowDuPont). In 2023, DuPont notified distributors that production of FOx-related products would cease at the end of 2025. To make nanofabrication routines for diamond quantum technologies future-proof, also in the light of cost-effectiveness (upper four digit \$ range per liter with minimum order quantity of one liter and shelf life of 6 months for FOx), we here present a process for a novel HSQ based resist called Medusa 84 SiH (Allresist GmbH), which has recently become commercially available. The active polymer HSQ is manufactured solely by Allresist and is not related to currently commercially available alternatives. In contrast to DuPont's FOx, a key difference lies in the chosen solvent which is butyl acetate instead of methyl isobutyl ketone (MIBK), where the latter is considered to be possibly carcinogenic to humans, meaning Medusa 84 SiH also comes with the advantage of higher personnel safety during its application. When stored at 10 $^\circ$C, its shelf life is analogous to DuPont's FOx (at least 3 months from the date of purchase, extending to 6 months) and deep-freeze storage to further prolong shelf life is possible. In this work, we use this novel resist to pattern diamond nanopillars as they are typical for scanning probe sensing using color centers as a prototypical nanostructuring process. The color centers embedded in our nanostructures are individual nitrogen vacancy (NV) centers, which are the most common color centers for quantum sensing.\\

\section{Nanofabrication Process}
\label{nanofab}

To make the following description of our nanofabrication process more readable, we summarize the parameters for devices, materials, and processes in tables in the supplementary material in table S.1 and S.2. In this study, we use three types of samples to evaluate the performance of Medusa 84 SiH. First, we use single crystal, single side polished silicon substrates (Microchemicals), to create accurate contrast curves and examine the writing dose-dependent etch resistivity of Medusa 84 SiH. Second, we use (100)-oriented, single crystal, chemical vapor deposited diamond (Element Six, 3.2 mm $\times$ 3.2 mm $\times$ (0.250 mm $\pm$ 0.050 mm)), for which the manufacturer specifies a moderate surface quality (R$_a$ $<$ 30 nm) and less than 1 ppm substitutional nitrogen [N]$_s$. We use this material to develop a consistent nanofabrication process and to study mask erosion. Third, we use a high purity (100)-oriented single crystal electronic grade diamond (Element Six), with dimensions of 4.0 mm $\times$ 2.0 mm $\times$ (0.047 mm $\pm$ 0.002 mm) ([N]$_s$ $<$ 5 ppb, [B] $<$ 1 ppb). This diamond has been polished by Delaware Diamond Knives to a surface roughness of R$_{q,\text{Before}}$ = 0.6 nm $\pm$ 0.5 nm as determined from 5 atomic force microscopy scans (5 $\mu$m $\times$ 5 $\mu$m, Bruker FastScan Bio). With this sample, we investigate how Medusa 84 SiH influences NV centers in nanopillars, which are often used in sensing applications.\\
We here follow published approaches to create shallow NV centers \cite{chu2014coherent, appel2016fabrication, radtke2019plasma}. To obtain coherent NV centers in low-strain material, we remove polishing damage prior to nitrogen implantation by conducting a "Stress Relief Etch" using an ICP-RIE (Oxford Instruments Plasmalab 100, for etch recipes refer to S.2) \cite{radtke2019plasma}. We implant nitrogen ions at 6 keV and a 7° angle to avoid ion channeling, resulting in a depth of 9.3 nm $\pm$ 3.6 nm as predicted by SRIM, at a fluence of 2 $\times$ 10$^{11}$ cm$^{-2}$. We anneal the diamond at 800 $^\circ$C for 2 hours under a vacuum of $<$ 7.8 $\times$ 10$^{-7}$ mbar using a custom-built furnace (Edwards T-Station 75, Tectra UHV compatible boralectric heater HTR-1001) to form NV centers \cite{chu2014coherent}. To remove non-diamond carbon that may form during annealing, we clean all diamonds in a Tri-Acid solution (HNO$_3$ 65 \%, HClO$_4$ 70 \%, H$_2$SO$_4$ 96 \% - 1:1:1) on a hot plate set at 500 $^\circ$C for 1 hour. We sonicate each sample sequentially in acetone (ACE) and isopropanol (IPA) before blow-drying with N$_2$. To improve handling, we glue the diamonds onto identical silicon substrates using a thin Crystal Bond 509 layer, followed by cleaning via stirring in IPA and blow-drying with N$_2$.\\
Processing diamonds using Medusa 84 SiH requires an adhesion promoting layer. Following our previous work on FOx 16 \cite{radtke2019reliable}, we deposit a 25 nm silicon layer using electron beam evaporation (Pfeiffer Classic 500 L). We note that thickness reduction of the layer might be feasible but has not been investigated in this work. This step was not used for the silicon substrates. To enhance adhesion of Medusa 84 SiH, we eliminate moisture through hotplate drying immediately prior to spin coating.\\
We apply Medusa 84 SiH (SX AR-N 8400, Allresist) using a spin coater (Süss Microtec RCD8). We reduce edge bead formation using a tailored two-step process, described below. Reducing edge bead formation here is crucial due to the small lateral dimensions of our diamonds. To gradually transfer the resist from the diamond to the silicon carrier chip, we increase rotational speed slowly to 1500 rpm (acceleration 500 rpm/s, 3 s at 1500 rpm). Subsequently, we ramp to 4000 rpm (acceleration 1000 rpm/s) and maintain 4000 rpm for 30 s. We developed this two-step recipe starting from our recipe for thicker resists such as FOx 16 (higher rotational speed ($\ge$ 5000 rpm) and acceleration of 2000 rpm/s in second step). Using the modified recipe lead to thicknesses ranging from 130 nm to 300 nm and patternable lateral space up to $\approx$ 47 \% of sample area. We apply the same recipe for all 3 sample types described above. We observe variability both between and within individual diamonds, attributed to lateral inhomogeneity caused by residual edge bead formation. Medusa 84 SiH underwent a soft bake on a hot plate set at 100 $^\circ$C for 2 minutes. In addition to the silicon adhesion promoting layer, which decreases charging during electron beam lithography \cite{radtke2019reliable}, we apply a layer of the conductive resist ESpacer 300Z (Showa Denko) at 4000 rpm on top of Medusa 84 SiH. This additional step minimizes electrical charging, thereby enhancing the effectiveness of proximity correction during electron beam lithography (EBL). We evaluated Electra 92 (Allresist) as alternative to ESpacer 300Z. Both performed similarly well when used with Medusa 84 SiH (see table S.1 for process details). We soft bake ESpacer 300Z at 80 $^\circ$C for 90 seconds. We note that Medusa 84 SiH is sensitive to both humidity and heat. Thus, all process steps after exposure to air were executed quickly to minimize effects on the effective writing dose. EBL was performed at 30 kV on a Raith eLiNE system (beam diameter $<$ 2 nm at 20 kV), utilizing focus-based working distance correction. We develop Medusa 84 SiH by stirring for 90 s in the developer AR 300-44 (Allresist), then rinsing with ultrapure water and cleaning with IPA for 30 s each, followed by blow-drying with N$_2$. We note as AR 300-44 is a developer based on 2.38 \% TMAH in water, the development also removes ESpacer 300Z or Electra 92. We report no residues originating from ESpacer 300Z or Electra 92 were observed during process testing. Both manufacturers suggest rinsing in ultra pure water (30 s - 60 s) to remove ESpacer 300Z / Electra 92 directly before resist development. We tested removing both during development and by rinsing prior to development without noticeable difference.\\
We use three designs for EBL. First, design 1 from which we obtain the contrast curve (60 $\mu$m squares and spacing). We expose each square uniformly with doses ranging from 50 $\mu$C/cm$^2$ to 5 mC/cm$^2$, adhering to a logarithmic sequence. Design 2 features multiple 25 $\mu$m squares and circles with diameters ranging from 16 $\mu$m to 180 nm, spacing varying between 3 $\mu$m - 6 $\mu$m (as shown in \textbf{Fig. \ref{fig:SEMImages} (a)}). We perform proximity correction using Raith's NanoPECS. This is accomplished via the dose scaling method. We employ the Monte Carlo simulation tool to estimate the proximity effect function, characterizing the radial energy distribution deposited by electrons in the resist. Most of the exposure dose results from low-energy secondary electrons generated by the high-energy electron beam during forward and backward scattering events within the resist. Utilizing the proximity function, the algorithm divides the initial design into smaller rectangles and basic polygons, such as triangles, allowing for adjustment with individual dose factors between 1 and 16 to achieve an approximately uniform dose distribution as intended by the original design. Design 3 consists of a 4 $\times$ 3 array of fields with 220 nanopillars (2.5 $\mu$m spacing) in each field and additional markers (letters, numbers, as shown in \textbf{Fig. \ref{fig:SEMImages} (b,c)}). Within the array, we adjust the nanopillar diameter (180 nm, 200 nm, 220 nm, and 240 nm) and an extra dose factor (0.9$\times$, 1.0$\times$, and 1.1$\times$). We thus enable the examination of NV center spin characteristics in different pillar geometries and partly counterbalance variations in the resist layer thickness due to edge beads. We subject design 3 to proximity correction with identical parameters, however, nanopillars were not segmented and written as circles with uniform doses. As a result from proximity correction, nanopillars typically received dose factors around 14 - 16 depending on their surroundings. For our studies, we use a dose array ranging from 160 $\mu$C/cm$^2$ to 430 $\mu$C/cm$^2$ in 30 $\mu$C/cm$^2$ increments. We here present the results achieved with a base dose of 310 $\mu$C/cm$^2$, therefore the nanopillars were manufactured typically with doses around the three nominal values: 4.464 mC/cm$^2$, 4.960 mC/cm$^2$, or 5.456 mC/cm$^2$.\\
To pattern the diamond, we employ two dedicated etching processes: Initially, we remove the silicon interlayer in-between the patterned mask using a biased SF$_6$ plasma. Subsequently, an oxygen-based "Pillar Etch" process transfers the mask pattern into diamond \cite{appel2016fabrication}. For both steps, we employ an ICP-RIE (Sentech PTSA-ICP Plasma Etcher SI 500, refer to table S.2 for parameters).\\
After structuring, we remove residual Medusa 84 SiH using buffered oxide etching (BOE) and residual silicon using potassium hydroxide (KOH). To finalize nanofabrication, we perform Tri-Acid cleaning to remove residual contamination and establish oxygen-containing surface termination followed by stirring in IPA and blow-drying with N$_2$ before characterizing the NV centers.\\

\section{Process Characterization}

This study reveals that Medusa 84 SiH offers performance comparable to the cutting-edge resist FOx, which is renowned for its effectiveness on diamond and similar substrates that require dense oxygen plasmas with significant physical etching in ICP-RIE, while allowing for nanoscale structuring. We have verified that charge dissipating resists like ESpacer 300Z or Electra 92 (Allresist) do not adversely affect Medusa 84 SiH. While an adhesion-promoting interlayer, such as silicon or titanium, is necessary for using Medusa 84 SiH, our results suggest that existing processes employing these with FOx by DuPont \cite{appel2016fabrication,radtke2019reliable, rani2020recent}, and potentially other HSQ variants by EM Resist, or Applied Quantum Materials, can be readily adapted to Medusa 84 SiH. With the discontinuation of DuPont's HSQ products, our research on Medusa 84 SiH supports it as a feasible substitute, filling the gap in various research domains that heavily relied on DuPont's FOx.\\

\subsection{Processing Quality}

\begin{figure}[h!]
	\centering
	\includegraphics[width=0.9\linewidth]{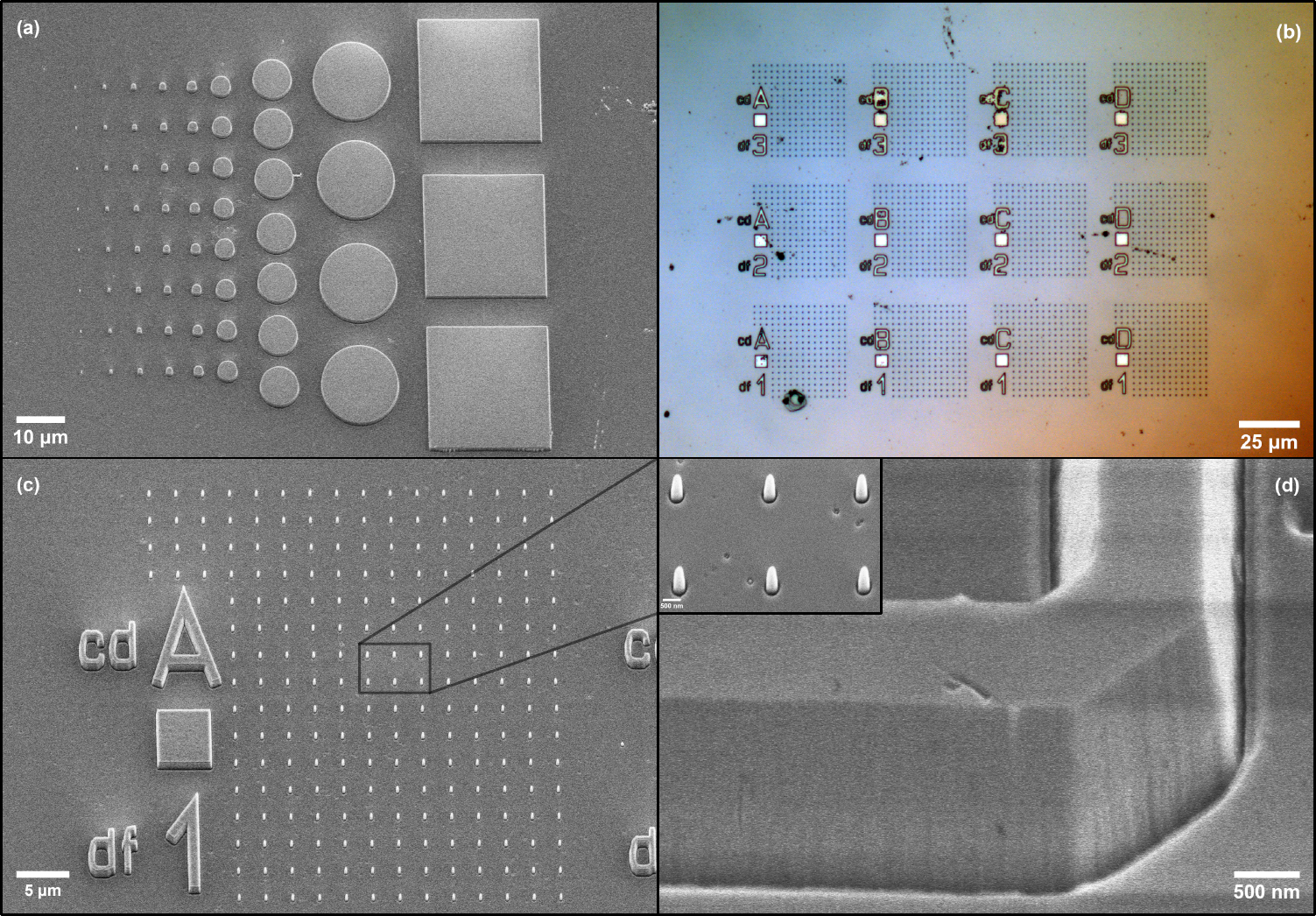}
	\caption{Scanning electron and optical microscopy images of diamond nanostructures fabricated using Medusa 84 SiH after ICP-RIE. (a-c) were imaged before mask removal, (d) after mask removal. All designs were proximity corrected using the commercial software NanoPECS by Raith GmbH. Nanopillars were not segmented during proximity correction. The images were captured at 5 kV acceleration voltage, at an 18.4 mm working distance and a 45$^\circ$ angle, with automatic tilt compensation enabled. (a) Dose test demonstrating the applicability of proximity correction to enhance corner and edge steepness of structures especially at higher spatial frequency. (b) Nanopillar field array for different nominal diameters (A: 180 nm, B: 200 nm, C: 220 nm, D: 240 nm) and resulting doses due to additional fine tuning dose factors df (1: 4.464 mC/cm$^2$, 2: 4.960 mC/cm$^2$, 3: 5.456 mC/cm$^2$) demonstrating high fabrication yield. Color change from "blue to red" from upper left to lower right corner is not connected to resist or processing. Surface contamination arises from several steps in our process (Tri-Acid treatment, ICP-RIE, manual sample handling) performed outside the cleanroom environment. (c) Nanopillar field written with a total dose of 4.464 mC/cm$^2$. (d) Sidewalls of a 915 nm high tall "B" indicator letter showing an intact top surface. Inset: Zoom-In of the A1 nanopillars from (c).}
	\label{fig:SEMImages}
\end{figure}

\textbf{Fig. \ref{fig:SEMImages} (a-c)} presents SEM images (Hitachi SU8000) and optical microscopy images of nanopillars etched into the electronic grade diamond before mask removal (4 min "Pillar Etch", pillar height 430 nm). \textbf{Fig. \ref{fig:SEMImages} (a)} illustrates micrometer-scale structures that have undergone proximity correction. The sharpness of corners and edges, the uniform thickness of the resist layer (data not shown) and the lack of both overexposure and thickness steps due to dose scaling evidence effective proximity correction, which is challenging on insulating substrates such as ultrapure diamond. \textbf{Fig. \ref{fig:SEMImages} (b)} presents the nanopillar array described above, indicating good adhesion properties to the silicon interlayer and therefore a high fabrication yield. Apart from small areas in which the adhesion-layer has been scratched, the fabrication yield of the nanopillars is only negligibly impacted by adhesion failure. \textbf{Fig. \ref{fig:SEMImages} (c)} presents the investigated nanopillar field with nominal diameters of 180 nm. As a result of proximity correction, even nanopillars close to larger marker structures and at the array border are written with defined, uniform shape. \textbf{Fig. \ref{fig:SEMImages} (d)} illustrates the sidewalls of a marker structure fabricated on a test diamond for examining mask erosion. The diamond was etched for 8 minutes with the "Pillar Etch", achieving a structure height of 915 nm $\pm$ 4 nm as determined by surface profilometry (Bruker DekTak XT, 1 \AA\ vertical resolution). The fabrication involved a Medusa 84 SiH layer, originally 300 nm $\pm$ 10 nm thick, which was thinned down by 90 nm $\pm$ 20 nm with 210 nm $\pm$ 10 nm remaining, demonstrating its suitability for O$_2$ plasma ICP-RIE processes (in-depth discussion in \textbf{Sec. \ref{Selectivity}}). The smooth and steep sidewall, displaying minimal roughness, indicates high writing quality, spatially homogeneous resist properties and minimal mask erosion. \textbf{Fig. \ref{fig:SEMImages} (d)} includes an inset providing a magnified view of six nanopillars with an aspect ratio of $\approx 2$ and a top diameter of $\approx$ 214 nm $\pm$ 5 nm indicating slight overexposure as lithography targeted $\approx$ 180 nm. SEM images indicate minimal roughening of the diamond surface as well as no mask redeposition effects related to Medusa 84 SiH. Utilizing an AFM (Park XE-70), we determine the surface roughness of the etched surface (from 12 scans, 1.25 $\times$ 1.25 $\mu$m). We find on average R$_{q,\text{After}}$ = 2 nm $\pm$ 1 nm slightly higher than R$_{q,\text{Before}}$ = 0.6 nm $\pm$ 0.5 nm. We identify individual etch pits as primary cause increasing R$_{q}$. As such individual etch pitch are not detrimental for the process demonstrated here, we excluded scans with prominent etch pits and determine R$_{q,\text{After}}$ = 0.8 nm $\pm$ 0.2 nm, close to R$_{q,\text{Before}}$. The findings clearly indicate that RIE-ICP etching utilizing Medusa 84 SiH maintains surface quality for well-polished substrates.\\

\subsection{Contrast}

\begin{figure}[h!]
\centering
	\includegraphics[width=1.0\linewidth]{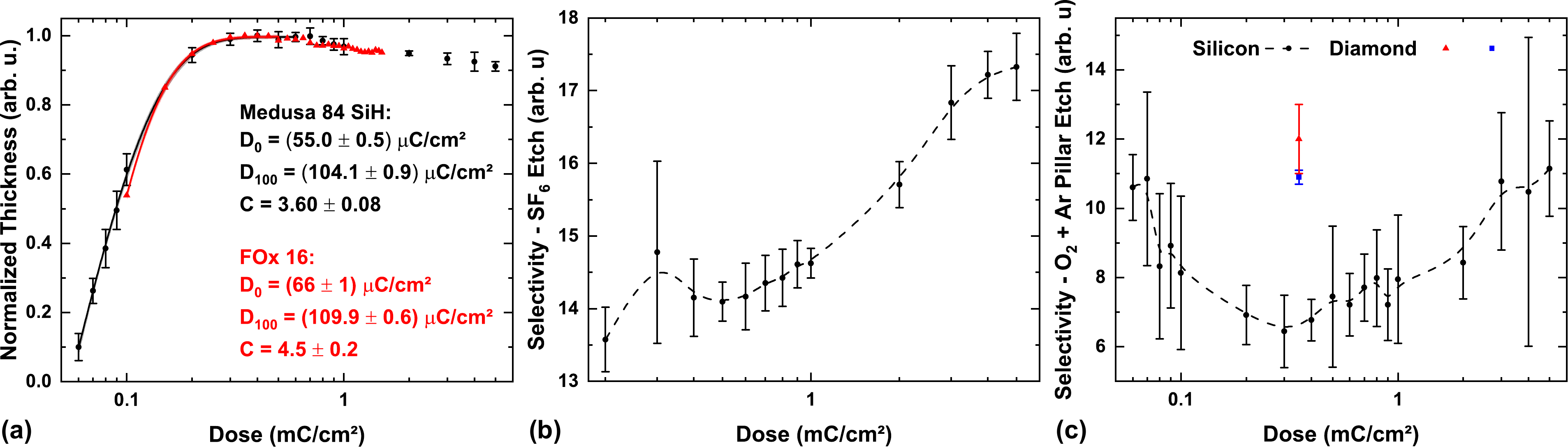}
	\caption{(a) Comparison of Medusa 84 SiH and FOx 16 contrast curves on silicon. To facilitate comparison, thickness values were normalized to their respective maxima: Medusa 84 SiH exhibited a layer thickness of 150.7 nm $\pm$ 0.3 nm, while the reference for FOx 16 was 819 nm $\pm$ 2 nm. For Medusa 84 SiH, each data point represents an average derived from nine identically processed duplicates. (b),(c) Dose-dependent etch selectivity for Medusa 84 SiH determined by evaluating ICP-RIE etched arrays of Medusa 84 SiH squares written with logarithmic dose progression. Each data-point corresponds to an average over nine identically processed duplicates. (b) Selectivity of adhesion layer removal etch, biased SF$_6$ etch, silicon : Medusa 84 SiH. (c) Selectivity of O$_2$ + Ar "Pillar Etch", diamond : Medusa 84 SiH. The dose-dependent Medusa 84 SiH etch rate was established from contrast curve patterns on silicon and compared against a reference diamond etch rate of 108 nm/min (see Table S.2). The dataset, obtained from test structures on two diamond samples with an optimal dose of 350 $\mu$C/cm$^2$ (colored), compares the realized etch rates for both Medusa 84 SiH and diamond. Red triangle: Medusa 84 SiH layer thickness of 130 nm $\pm$ 20 nm, etched for 6 min. Blue square: Medusa 84 SiH layer thickness of 300 nm $\pm$ 10 nm, etched for 8 min.}
	\label{fig:FabCurves}
\end{figure}

We characterize the resist's writing dose and contrast by evaluating nine identically written Medusa 84 SiH arrays of squares with logarithmic dose progression on a single silicon substrate and averaging their dose-dependent residual thicknesses post-development. For comparison with previous FOx 16 contrast curves, the contrast $C$ is calculated using the standard definition from \cite{campbell2008fabrication}:

\[C = \frac{1}{\text{Log}_{10}(D_{100} / D_0)} \text{\hspace{0.2cm}.}\]

Following this definition \cite{campbell2008fabrication}, $D_{100}$ denotes the base dose that separates grayscale from the conventional lithography regions, while $D_0$ represents the minimum dose required for structures to appear. To find $D_0$, we utilize this fitting function:

\[T(D) = T_0 (1 - \text{exp}(-\frac{D-D_0}{D_c})) \text{\hspace{0.2cm}.}\]

$T$ is the resist thickness, $D$ denotes the dose, $T_0$ is the layer thickness for full exposure, and $D_c$ is a parameter that accounts for the exponential's slope and consequently the contrast $C$. To find $D_{100}$, we calculate where the horizontal line $T(D) = T_0$ intersects the tangent to the fit function at $D_0$. \textbf{Fig. \ref{fig:FabCurves} (a)} shows the measured contrast curves. For better comparison, we normalize both contrast curves to their respective maxima. On average, using the spin coating recipe described above, we achieve with Medusa 84 SiH a layer thickness of $T_0$ = 150.7 nm $\pm$ 0.3 nm on silicon at a maximum rotational speed of 4000 rpm. The initial 60 $\mu$m squares emerge above $D_0$ = 55.0 $\mu$C/cm$^2$ $\pm$ 0.5 $\mu$C/cm$^2$, and we identified a base dose of $D_{100}$ = 104.1 $\mu$C/cm$^2$ $\pm$ 0.9 $\mu$C/cm$^2$. We find $C_{\text{Medusa}}$ = 3.6 $\pm$ 0.08, similar to FOx 16 $C_{\text{FOx}}$ = 4.5 $\pm$ 0.2 under similar processing conditions in our previous work \cite{radtke2019plasma, radtke2019reliable, appel2016fabrication, neu2014photonic}. We note that the FOx 16 layer was almost 700 nm thicker and we used a different developer so the results are only partially comparable.\\

\subsection{Selectivity}
\label{Selectivity}

Furthermore, we examine if the etch resistance of Medusa 84 SiH depends on the dose. We note that for thick resist layers, the selectivity might vary with thickness because of depth-dependent effective dose distribution within the layer. Consequently, thickness variations might influence the evaluation of selectivity especially for diamond where thickness varies stronger. We expose our above discussed contrast curve pattern for 1 minute to the "Pillar Etch" plasma and another pattern on a second silicon substrate for 45 s to the biased SF$_6$ plasma used for silicon adhesion layer removal. To determine the selectivity, we use the following approach. We first measure the surface profile of the sample after ICP-RIE via profilometry. We then remove the Medusa 84 SiH mask using BOE treatment which induces negligible etching of the silicon substrate. We again measure the surface profile of the silicon substrate. We now extract the resist thickness via comparing the surface profiles to account for etching of the silicon substrate in the plasma. Utilizing the resist's thicknesses before and after ICP-RIE, we compute the etch rates of the plasmas for the materials involved and determine selectivity ratios. \textbf{Fig. \ref{fig:FabCurves} (b,c)} summarizes the resulting, dose-dependent selectivities (details see caption). To estimate selectivity on diamond, we repeat evaluation for each ICP-RIE etched diamond using three randomly selected, but neighboring 25 $\mu$m square structures for each diamond for the base dose for which we observed optimal writing quality. We note that even for these neighboring structures, we observe thickness variations. On diamond, we observe a decreased etch rate of the "Pillar Etch" plasma on Medusa 84 SiH compared to silicon leading to increased selectivities for the chosen dose of 350 $\mu$C/cm$^2$. For a diamond featuring a Medusa 84 SiH layer of 130 nm $\pm$ 20 nm thickness, etched for 6 minutes, the selectivity was found to be 12 $\pm$ 1. For a second diamond with higher layer thickness of 300 nm $\pm$ 10 nm, etched for 8 minutes we find a selectivity of 10.9 $\pm$ 0.2. Both indicate that the insulating diamond substrate influences plasma dynamics and reduces the etch rate of the "Pillar Etch" on Medusa 84 SiH by approximately 40 \%, favoring its application in diamond nanofabrication. Moreover, beyond $D_{100}$, there is a notable increase in selectivity. We suspect enhanced cross-linking and resist densification to cause this effect, indicating that further hard baking post-development could improve etch resistance at lower lithography dose. The selectivity values presented in this study are comparable to those previously reported for similar ICP-RIE recipes using HSQ or SiO$_2$-like masks \cite{radtke2019reliable,hausmann2010fabrication}.\\

\subsection{Fabrication Yield}

We analyze optical microscopy images of all diamond nanopillars fields to quantify the fabrication yield of nanopillars. We note that prior to this evaluation, we used SEM imaging to make sure the structures we observe in optical microscopy are intact nanopillars and not structures created due to collapsed pillar masks that adhered to their original positions. A total of 79.200 nanopillars have been evaluated on a single electronic grade diamond. We report a total fabrication yield of 89.6 \%. Among the 10.4 \% fabrication failures, 59.4 \% result from a non-optimal synthesis and oligomeric silsesquioxanes \cite{frye1970oligomeric}, 20.7 \% are caused by edge beads that affect lithography and 17.7 \% are the result of adhesion failure. The remaining 2.2 \% is due to contaminants, as well as overexposure in lithography. The work presented here is based on an early prototype of Medusa 84 SiH by Allresist. The now commercially available version has been significantly optimized regarding manufacturing to remove the effect originating from non-optimal synthesis. We note original FOx resist was also prone to similar issues before synthesis optimization \cite{frye1970oligomeric}. Therefore fabrication yields could potentially reach around 96 \%, primarily constrained by edge beads on small diamond substrates and sporadic adhesion failures.\\

\section{Experimental Setup}

We used a custom-built confocal microscope (modified version of \cite{no2022}) to study the impact of diamond nanostructuring with Medusa 84 SiH on NV centers by comparing their room-temperature spin properties before and after processing. As our goal is to obtain pillars with a high probability of finding single NV, the NVs in the implanted layer are so dense, that we observe small ensembles in our confocal microscope instead of individual NV centers prior to nanopillar fabrication. We utilize a diode laser at 520 nm, featuring sub-nanosecond digital modulation (DL nSec, PE 520, Swabian Instruments) and a maximum output power of 40 mW to excite and initialize the NV centers. A 600 nm shortpass filter placed behind the single-mode fiber guiding the excitation light eliminates background light from the fiber. The microscope objective (LMPLFLN100X, Olympus) has a numerical aperture of 0.8, along with a working distance of 3.4 mm. We utilize a piezo-electric scanner (ANSxyz100, standard range, Attocube Systems) to facilitate sample scanning. We use two dichroic mirrors and a 600 nm longpass filter to separate excitation and fluorescence light. We sent the fluorescence via a single mode fiber to a fiber coupled avalanche photo diode (APD, SPCM-AQRH-14-FC, Excelitas Technologies) with a quantum efficiency of approximately 69 \% at 700 nm. A data acquisition card (PCIe-6323, National Instruments) acquires and logs the signal. To analyze fluorescence spectra, we direct the light to a diffraction grating spectrometer (Acton SpectraPro SP-2500, Princeton Instruments) that records the spectra using a CCD camera (Pixis 256E, Princeton Instruments) using a multi-mode fiber replacing the single mode detection fiber. To investigate excited state lifetimes via time correlated single photon counting, we use a supercontinuum laser (SuperK Extreme EXW-12, NKT-Photonics) with a configurable single-line filter (SuperK VARIA, NKT-Photonics) and a time-tagging module (PicoHarp 300, PicoQuant). For coherent spin manipulation, a microwave source (TSG4100A, Tektronix) is connected to a +45 dB amplifier (ZHL-16W-43+, Mini-Circuits), feeding an$\Omega$-shaped stripline microwave antenna \cite{opaluch2021optimized}, on which the diamond is fixed using Crystal Bond 509 (Structure Probe Inc). We apply adjustable external magnetic fields using neodymium magnets on a rotation mount and a three-axis linear translation stage.\\

\section{Shallow NV Centers in Nanopillars manufactured using Medusa 84 SiH Masks}

\begin{figure}[h!]
	\centering
	\includegraphics[width=0.9\linewidth]{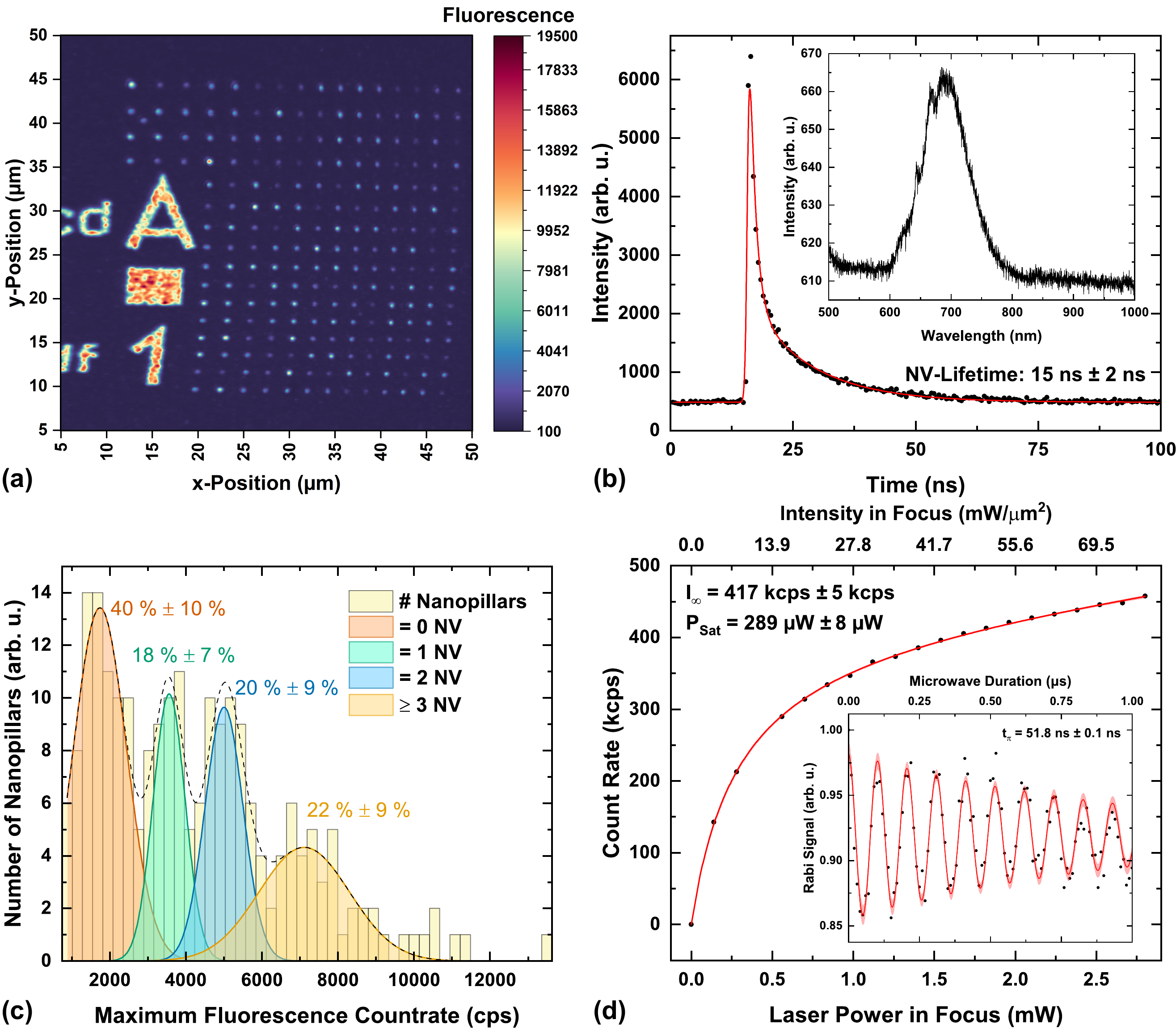}
	\caption{Fluorescence microscopy of diamond nanopillars with typically 0 - 3 NV centers fabricated using Medusa 84 SiH. (a) Confocal microscopy xy-scan of a nanopillar field (diameters $\approx$ 214 nm $\pm$ 5 nm). (b) Excited state lifetime curve of a single NV-Center within a nanopillar with a value of $\tau_{\text{NV}}=$ 15 ns $\pm$ 2 ns. Fit was done using a bi-exponential decay convoluted with a Gaussian to account for the setup's instrument response function (244 ps width). Bi-exponential decay fit was necessary due to the presence of short living surface fluorescence ($\tau_{\text{Surface}}=$ 1.9 ns $\pm$ 0.1 ns, refer \cite{xiao2015fluorescence,padrez2023nanodiamond}). Inset: Fluorescence spectrum of a single nanopillar showing a dominant NV$^-$ spectrum, the NV$^-$ ZPL at 637 nm is also visible. (c) Histogram displaying the count of nanopillars categorized by their peak fluorescence count rate. Based on fluorescence intensity, the estimated distribution of NV centers per nanopillar is as follows: 0 NV: 40 \% $\pm$ 10 \%, 1 NV: 18 \% $\pm$ 7 \%, 2 NV: 20 \% $\pm$ 9 \%, 3 NV: 22 \% $\pm$ 9 \%. (d) Saturation curve illustrating the photonic enhancement in the fluorescence emission of a single NV center relative to bulk. With a reference saturation count rate of $I_{\infty\text{,Bulk}}$ = 60 kcps $\pm$ 7 kcps, this results in a 7.0 $\pm$ 0.9 increase in photon collection efficiency. The shown count rates were corrected for the non-linear response of our APD following the manufacturer's guidelines. The fitting function used incorporates a power-dependent background term. Inset: A typical Rabi oscillation with a $\pi$-pulse duration of t$_\pi$ = 51.8 ns $\pm$ 0.1 ns as realized on nanopillars with single NV centers.}
	\label{fig:FluorescenceMicroscopy}
\end{figure}

\textbf{Fig. \ref{fig:FluorescenceMicroscopy} (a)} illustrates a confocal xy-scan of a nanopillar array (diameters $\approx$ 214 nm $\pm$ 5 nm). The fluorescence intensity emitted from the nanopillars is typical for individual NV centers in such structures \cite{nelz2019toward,appel2016fabrication,neu2014photonic,babinec2010diamond,hausmann2010fabrication}. It should be noted that all the measurement data displayed graphically stem from nanopillars exhibiting single NV centers. This was checked by their brightness and the presence of maximally one resonance pair during optically detected magnetic resonance (ODMR) within differently oriented external magnetic fields. Fluorescence spectra of pillars reveal the spectrum of NV centers, characterized by their broad phononic sideband and a zero phonon line at 637 nm, as shown in the inset of \textbf{Fig. \ref{fig:FluorescenceMicroscopy} (b)}. We perform lifetime measurements and fit them using a bi-exponential decay convoluted with a Gaussian to account for the instrument response function of our setup (244 ps width) as shown in \textbf{Fig. \ref{fig:FluorescenceMicroscopy} (b)}. Using the bi-exponential decay is necessary as we observe a fast decaying fluorescence component that we attribute to (broadband) surface fluorescence ($\tau_{\text{Surface}}=$ 1.9 ns $\pm$ 0.1 ns) \cite{xiao2015fluorescence,padrez2023nanodiamond} which does not hinder using these structures in quantum sensing. We determine the NV excited state lifetime to be $\tau_{\text{NV}}=$ 15 ns $\pm$ 2 ns, which is slightly increased in comparison to NV centers in bulk diamond due to their proximity to the surface and excludes excessive quenching. To estimate the number of NV centers per nanopillar, we evaluated xy-scans for nanopillar fields. By identifying the local maxima of the histogram in \textbf{Fig. \ref{fig:FluorescenceMicroscopy} (c)} as occurring due to the presence of 0, 1, 2, or 3 NV centers and assuming Gaussian distributions, we approximated the occurrence of a single NV center per nanopillar as 18 \% $\pm$ 7 \% (details see caption) comparable to Ref.\ \cite{appel2016fabrication}. Using this distribution, we estimate the creation yield of shallow NV centers in our sample. We determined an NV center density of \(3 \times 10^9 \, \text{cm}^{-2} \pm 2 \times 10^9 \, \text{cm}^{-2}\), indicating a creation yield of 1.7 \% $\pm$ 0.8 \% which is comparable to other works \cite{appel2016fabrication} excluding excessive charge-state instability. We note that this approach only roughly estimates the NV density as the expected photonic enhancement and thus the apparent brightness of NV centers in nanopillars has been found to strongly depend on the spatial position within the nanopillar (see, e.g. Ref. \cite{neu2014photonic}). We note that the observed NV centers are stable and do not show blinking or bleaching. This observation together with the NV creation yield matching previous work indicate that Medusa 84 SiH efficiently protects shallow NV centers during ICP-RIE.\\

\textbf{Fig. \ref{fig:FluorescenceMicroscopy} (d)} highlights a saturation measurement of a pillar containing a single NV center. The displayed count rates have been corrected according to the non-linear response of our APD, as per manufacturer's instructions. We applied the given equation to determine the saturation power and fluorescence count rate of the NV center \cite{neu2014photonic}:

\[I(P) = I_\infty \frac{P}{P+P_{\text{Sat}}} + b_g P \text{\hspace{0.2cm}.}\]

Here, $I$ is the observed fluorescence count rate, $P$ denotes the laser power after the microscope objective, $I_\infty$ represents the saturated fluorescence count rate, $P_\text{Sat}$ is the saturation power, and $b_g$ is the proportionality factor for the background, scaling linearly with laser power. Comparison between the saturation count rate of a single NV center within a nanopillar $I_{\infty}$ = 417 kcps $\pm$ 5 kcps in this work and results in bulk $I_{\infty\text{,Bulk}}$ = 60 kcps $\pm$ 7 kcps from our previous work \cite{no2022}, results in a 7.0 $\pm$ 0.9 increase in photon collection efficiency, as a result of efficient wave-guiding via the nanopillar's photonic modes, which is at the same order of magnitude of previously reported values \cite{babinec2010diamond,appel2016fabrication}.\\

\begin{figure}[h!]
	\centering
	\includegraphics[width=0.9\linewidth]{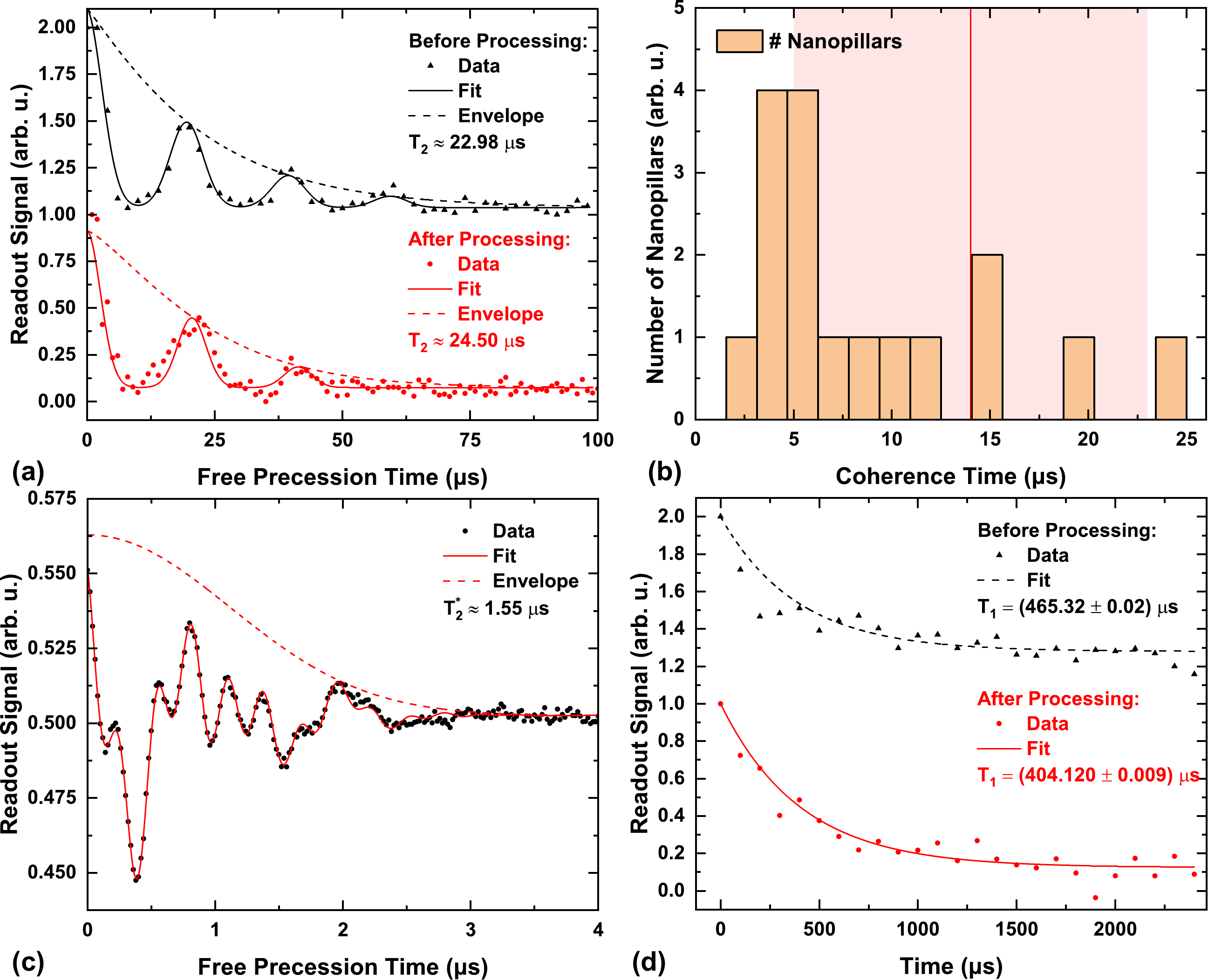}
	\caption{Investigations comparing shallow NV center spin properties before processing and within nanopillars processed using Medusa 84 SiH. Where applicable, readout signals have been normalized with their maximum value and shifted by one for better comparison. (a) Spin Echo measurements, similar maximum values for coherence times T$_2 \approx$ 24 $\mu$s were found before and after processing. (b) Histogram showing distribution of coherence times T$_{2,\text{After}}$ of NV center after processing. In red average and standard deviation of T$_{2,\text{Before}}$ before processing. A potential decrease in coherence time seems to be negligible. (c) Ramsey measurement showing a dephasing time of T$^*_2 \approx 1.55$ $\mu$s in line with values typically found in commercial NV scanning probes \cite{Qnami2021}. (d) Spin lifetime measurements showing comparable T$_1$ values around 430 $\mu$s before and after processing.}
	\label{fig:SpinProperties}
\end{figure}

We examine spin characteristics of the shallow NV ensemble before processing and for individual NV in nanopillars after processing to assess their usability as quantum sensors. To this end, we perform optically detected magnetic resonance in presence of an external magnetic field. After processing, we select nanopillars for further investigation with single NV centers. We select a single transition for resonant driving of Rabi oscillations, to determine the $\pi$-pulse duration as shown in the inset of \textbf{Fig. \ref{fig:FluorescenceMicroscopy} (d)}. Subsequently, we perform spin echo experiments to determine the NV center's coherence time $T_2$, as illustrated in \textbf{Fig. \ref{fig:SpinProperties} (a)}. This procedure was repeated for 17 nanopillars, each containing a single NV center, we compare these $T_2$ values to $T_2$ of the shallow NV ensemble before nanopillar fabrication. We employ a fitting function from \cite{no2022} to extract $T_2$:

\[ I(\tau) = \frac{\alpha_0 + \alpha_1}{2} + \frac{\alpha_0 - \alpha_1}{2} \exp\left[ -\left( \frac{\tau}{T_2}\right) \right]^n \sum\limits_{0}^j \exp\left[ -\left( \frac{\tau - j \tau_\text{Rev}}{T_\text{Dec}}\right)^2\right] \cos\left(2 \pi \gamma B_\text{NV} t\right).\]

In this context, $\alpha_0$ and $\alpha_1$ represent the fluorescence count rates for the bright state $\left| 0 \right\rangle$ and dark state $\left| \pm 1 \right\rangle$, respectively. Here, $\tau$ denotes the free evolution duration, $n$ is the associated order of the decoherence mechanism, $j$ is the count of observed spin revivals, $\tau_\text{Rev}$ is the revival period of the Larmor precession, $T_\text{Dec}$ the correlation time of the surrounding $^{13}$C spin bath, $\gamma$ is the gyromagnetic ratio of the NV center's electronic spin and B$_\text{NV}$ the magnetic field at its location.\\
In this study, we found an average coherence time of $T_{2\text{,Before}}$ = 14 $\mu$s $\pm$ 9 $\mu$s before nanopillar fabrication using Medusa 84 SiH and $T_{2\text{,After}}$ = 9 $\mu$s $\pm$ 6 $\mu$s after processing. As shown in \textbf{Fig. \ref{fig:SpinProperties} (b)}, the coherence times are in the same window as observed before, which indicates that processing diamond using Medusa 84 SiH does not strongly reduce shallow NV center's coherence times T$_2$. The determined coherence times $T_2$ are in the same order of magnitude as those reported for single NV centers within nanopillars fabricated using FOx, where nanofabrication was also found to reduce the average $T_2$ \cite{appel2016fabrication}. Additionally, Ramsey measurements, as displayed in \textbf{Fig. \ref{fig:SpinProperties} (c)}, have been conducted. In average, we found dephasing times of $T^*_2$ = 1.2 $\mu$s $\pm$ 0.4 $\mu$s which lies in the same range as for commercially available single NV scanning probes \cite{Qnami2021}. Moreover, T$_1$ spin lifetime measurements were performed before and after processing. As shown in \textbf{Fig. \ref{fig:SpinProperties} (d)}, the spin lifetime also shows only a slight decrease after processing (average T$_1$ = 0.3 ms $\pm$ 0.1 ms ). While T$_1$ before and after processing is consistent for our diamond sample, we note that we typically find T$_1$ times of several milliseconds for commercially available NV scanning probes \cite{Qnami2021}. These results conclusively demonstrate that the use of Medusa 84 SiH for diamond nanofabrication, compared to the cutting-edge HSQ resist FOx by DuPont, does not adversely affect the spin properties of color centers, thus rendering Medusa 84 SiH appropriate for diamond-based quantum technologies.\\

\section{Conclusion}

This study examined Allresist's novel electron beam resist Medusa 84 SiH as a substitute for DuPont's Flowable Oxide for diamond nanofabrication, focusing on its use in diamond-based quantum technologies. We transferred previously established processes for FOx to Medusa 84 SiH \cite{appel2016fabrication,neu2014photonic,radtke2019reliable} and proved their feasibility by realizing prototypical test structures at the $\mu$m-scale down to nanopillars with diameters around $\approx$ 200 nm. We investigated the resist's basic properties by evaluating fabrication yield, recording contrast curves and determining its dose-dependent etch resistance against common plasmas necessary to nanostructure diamond. We confirm that Medusa 84 SiH is very similar to FOx for the demonstrated prototypical structures with comparable contrast, achieved fabrication yields of up to $\approx$ 96 \% for electron beam evaporated silicon as adhesion promoting interlayer and shows a conservatively estimated selectivity towards high biased O$_2$ + Ar ICP-RIE plasma in diamond etching of 11 to 12 for $\mu$m-scale structures. We further investigated the impact of processing diamond with Medusa 84 SiH on shallow NV centers within nanopillars as applied in the field of quantum sensing \cite{appel2016fabrication,neu2014photonic,Qnami2021} by comparing T$_2$, T$^*_2$ and T$_1$ times before and after fabricating nanopillars. We report negligible decreases in spin lifetime, dephasing and coherence times, therefore proving the suitability of these nanopillars for quantum sensing schemes. Further, we investigated their fluorescence emission and show that, with the proposed nanofabrication process, an enhancement in photon collection efficiency of 7.0 $\pm$ 0.9 was achieved. Overall, Medusa 84 SiH is a promising alternative to FOx as an electron beam resist for diamond nanofabrication compatible with quantum technologies like NV sensing. Future optimization might include contrast optimization, e.g. lowering the exposure to thermal energy during soft baking and adjusting the development parameters. In addition to diamond, this work hints Medusa 84 SiH might be also an appropriate alternative for FOx in other material systems like silicon, SOI, GaAs or other materials under research requiring nanoscale resolution and high resistance towards RIE processes.\\

\section{Supplementary Materials}

\subsection{Nanofabrication}

\begin{table}[h!]
\caption{Medusa 84 SiH Nanofabrication Process}\label{ProcessParameters}
  \begin{tabular*}{\textwidth}{@{\extracolsep\fill}ll}
    \toprule
    \textbf{Process Step} & \textbf{Process Parameters} \\
    \midrule
  Substrate Characteristics & Single crystalline, electronic grade, (100)-oriented CVD diamond\\
     & Surface Roughness: 0.6 nm $\pm$ 0.5 nm\\
     & Size: 4 mm $\times$ 2 mm $\times$ (0.047 mm $\pm$ 0.002 mm)\\
     & Wedge Dimensions:\\
     & \hspace{15pt} Long Side: 2.6 $\mu$m $\pm$ 0.6 $\mu$m, Short Side: 2.2 $\mu$m $\pm$ 0.7 $\mu$m\\
     & Impurity Levels: [N]$_s$ $<$ 5 ppb, [B] $<$ 1 ppb\\
     & $^{13}$C Proportion: 1.109\%\\
     & Suppliers:\\
     & \hspace{15pt} Synthesis: Element Six, Polish: Delaware Diamond Knives\\
    \midrule
    Tri-Acid & 65\% HNO$_3$, 70\% HClO$_4$, 96\% H$_2$SO$_4$ - 1:1:1 at 500 $^\circ$C for 1 h\\
    \midrule
    Ultrasonic Bath & ACE $\Rightarrow$ IPA, 6 min at room temperature\\
    \midrule
    Stress Relief Etch & Deep Etch: 3.66 $\mu$m, Biased Etch: 100 nm, 0V Bias Etch: 4 nm\\
    & See \textbf{Table \ref{EtchRecipes}} for plasma parameters\\
    \midrule
    Ion Implantation & 2 $\times$ 10$^{11}$ cm$^{-2}$ N$^+$ at 6 keV with angle of 7$^\circ$, 9.3 nm $\pm$ 3.6 nm\\
    \midrule
    High Vacuum Annealing & Heating at 800 $^\circ$C for 2 h starting below 7.8 $\times$ 10$^{-7}$ mbar\\
    \midrule
    Substrate Preparations & Tri-Acid and Ultrasonic Bath, see the preceding text\\
    \midrule
    Electron Beam Evaporation & 25 nm Silicon deposited at 4.5 \AA/s under 3.0 $\times$ 10$^{-6}$ mbar\\
    \midrule
    Moisture Removal & 100 $^\circ$C for 6 min\\
    \midrule
    Spin Coating of Resist Layer & Step \#1: 1500 rpm for 6 s, Acc.: 500 rpm/s\\
     & Step \#2: 4000 rpm for 30 s, Acc.: 1000 rpm/s\\
     & Resist: SX AR-N 8400 (Medusa 84 SiH)\\
     & Thickness: 150 nm, Soft Bake: 100 $^\circ$C for 2 min\\
    \midrule
    Application of conductive Layer & Resist: ESpacer 300Z, 4000 rpm for 30 s, Acc.: 1000 rpm/s\\
     & Layer Thickness: 20 nm, Soft Bake: 80 $^\circ$C for 90 s\\
     & Resist: Electra 92, 2000 rpm for 60 s, Acc.: 1000 rpm/s\\
     & Layer Thickness: 60 nm, Soft Bake: 90 $^\circ$C for 120 s\\
    \midrule
    Proximity Correction & $\frac{1}{\pi (1 + \eta + \nu_1 + \nu_2)}(\frac{1}{\alpha^2}e^{-\frac{x^2}{\alpha^2}} + \frac{\eta}{\beta^2}e^{-\frac{x^2}{\beta^2}} + \frac{\nu_1}{2 \gamma_1^2}e^{-\frac{x}{\gamma_1}} + \frac{\nu_2}{\gamma_2^2}e^{-\frac{x^2}{\gamma_2^2}})$\\
     & $\alpha$ = 4 nm, $\beta$ = 8816 nm, $\gamma_1$ = 126 nm, $\gamma_2$ = 237 nm\\
     & $\eta$ = 23.393483, $\nu_1$ = 1.56053, $\nu_2$ = 1.015\\
    \midrule
    Electron Beam Lithography & Acceleration Voltage: 30 kV, Aperture: 30 µm\\
     & Working Distance: 6.107 mm, Step Size: 12.8 nm\\
     & Dose Arrray: 160 $\mu$C/cm$^2$ - 430 $\mu$C/cm$^2$\\
     & Working Distance Correction: Focus\\
    \midrule
    Development & AR 300-44 (2.38\% TMAH) for 90 s\\
     & Rinse: Ultrapure water for 30 s, Cleaning: Isopropanol for 30 s\\
    \midrule
    Structuring & Biased SF$_6$: 21 nm, Pillar Etch: 432 nm\\
     & See \textbf{Table \ref{EtchRecipes}} for plasma parameters\\
    \midrule
    Mask Removal & Buffered Oxide Etch 12.5\% for 20 minutes at room temperature\\
    \midrule
    Adhesion Layer Removal & 3 M Potassium Hydroxide at 80 $^\circ$C for 30 min\\
    \midrule
    Final Cleaning & Tri-Acid, see the preceding text\\
     & Stir for 30 s in ACE $\Rightarrow$ IPA at room temperature\\
    \botrule
  \end{tabular*}
\end{table}

\textbf{Table \ref{ProcessParameters}} provides a comprehensive summary of all process steps in sequential order, detailing the majority of the process parameters specifically related to the electronic grade diamond. For diamonds of lower purity utilized in process development, the stress relief etch, ion implantation and high vacuum annealing steps were omitted. Moreover, the silicon samples used for contrast curve analysis did not undergo Tri-Acid or KOH treatments, and procedures such as electron beam evaporation, deposition of a conductive layer, and proximity correction were also excluded.\\

\begin{table}[h!]
\caption{ICP-RIE Etch Recipes}\label{EtchRecipes}
\begin{tabular*}{\textwidth}{@{\extracolsep\fill}lcccccccc}
\toprule
& \multicolumn{3}{@{}c@{}}{Gas Composition [sccm]} & \multicolumn{2}{c@{}}{Power [W]} \\\cmidrule{2-4}\cmidrule{5-6}
\midrule
\textbf{Stress Relief\footnotemark[1]} & O$_2$ & Ar & SF$_6$ & RF & ICP & DC Bias & Pressure & Etch Rate\footnotemark[3] \\
\midrule
Deep Etch & 22.5 & 15 & 7.5 & 100 & 700 & 300 V & 12 mTorr & 61 nm/min\\
Biased O$_2$ Etch & 50 & 0 & 0 & 165 & 500 & 450 V & 11 mTorr & 98 nm/min\\
0V DC Bias Etch & 50 & 0 & 0 & 0 & 550 & 0 V & 11 mTorr & 8 nm/min\\
\midrule
\textbf{Structuring\footnotemark[2]} & O$_2$ & Ar & SF$_6$ & RF & ICP & DC Bias & Pressure & Etch Rate\footnotemark[3] \\
\midrule
Biased SF$_6$ Etch & 0 & 0 & 50 & 60 & 500 & -160 V & 1 Pa & 2.647 $\mu$m/min\footnotemark[4]\\
Pillar Etch & 30 & 25 & 0 & 100 & 450 & -167 V & 1 Pa & 108 nm/min\\
\botrule
\end{tabular*}
\footnotetext[1]{Were performed on Oxford Instruments Plasmalab 100.}
\footnotetext[2]{Were performed on Sentech PTSA-ICP Plasma Etcher SI 500.}
\footnotetext[3]{Were determined via surface profilometry using a Bruker Dektak XT.}
\footnotetext[4]{Etch rate for silicon.}
\end{table}

\textbf{Table \ref{EtchRecipes}} summarizes the parameters used to establish the ICP-RIE plasmas used to conduct stress relief etches on an Oxford Instruments Plasmalab 100 or to structure diamond with the proposed fabrication process using a Sentech PTSA-ICP Plasma Etcher SI 500. It should be noted that both ICP-RIE devices can realize the described processes, although their parameters require appropriate adjustment.\\

\backmatter

%\bmhead{Supplementary information}

\bmhead{Acknowledgements}

We thank the Nano Structuring Center at RPTU Kaiserslautern-Landau (Kaiserslautern, Germany), especially Dr. Bert Lägel for assisting with electron beam lithography and Medusa 82 prototyping, and Alexander Frick for silicon wafer cutting and maintenance. We also thank the Leibniz Institute for New Materials (Saarbrücken, Germany), particularly Dr. René Hensel for use of their Oxford Instruments Plasmalab 100. We thank Dr. Matthias Schreck and Wolfgang Brückner for performing nitrogen implantation. The authors also acknowledge Kilian Mark and Günter Marchand (Universität des Saarlandes, Saarbrücken, Germany) for their help with the Sentech PTSA-ICP Plasma Etcher SI 500. The authors thank Dr. Dipti Rani and Dr. Lahcene Mehmel (LSPM, Villetaneuse, France) for early contributions to this project. We acknowledge Yanis Abdedou and Philipp Fischborn (Workgroup E. Neu) for technical support, supporting fabrication and optical characterization and fruitful discussions. Yves Kurek assisted with SRIM simulations.\\

\section*{Declarations}

\textbf{Funding:} Funding for this work was provided by the Deutsche Forschungsgemeinschaft (DFG, German Research Foundation) under Grant No. TRR 173–268565370, Spin+X (Project A12) and TRR 306-429529648, QuCoLiMa and DFG: INST 256/533-1 FUGG . EN acknowledges support from the Quantum-Initiative Rhineland-Palatinate (QUIP).\\
\textbf{Conflict of Interest:} Harry Biller and Mandy Sendel are employees of Allresist GmbH, the vendor of the Medusa 84 SiH resist used in this study. Allresist has supplied the pre-commercial prototype for these investigations free of charge. All other authors certify that they have no affiliations with or involvement in any organization or entity with any financial interest or non-financial interest in the subject matter or materials discussed in this manuscript.\\
\textbf{Data Availability:} The data used in this manuscript has been made publicly available via the Zenodo repository: https://doi.org/10.5281/zenodo.14976451\\
\textbf{Author Contributions:} E.Neu and C. Becher obtained the funding while E.Neu, O.R.Opaluch, S.Wolff, H. Biller, M. Sendel and C.Becher planned the study. E.Neu, O.R.Opaluch, S.Wolff, H. Biller and M. Sendel devised the fabrication method. Investigations were planned by E.Neu, O.R.Opaluch and N.Oshnik, who also oversaw the research. O.R.Opaluch executed the nanofabrication along with its characterization and data analysis. The Sentech PTSA-ICP Plasma Etcher SI 500 procedures were developed by P.Fuchs and J.Fait. NV Sensing experiments, along with characterization and analysis, were performed by O.R.Opaluch, S.Westrich and N.Oshnik. O.R.Opaluch prepared the initial manuscript draft, and all authors engaged in its review and editing.\\

%\noindent

%\begin{appendices}

%\end{appendices}

%\bibliographystyle{sn-mathphys-num}
\bibliography{sn-bibliography}

%% BioMed_Central_Bib_Style_v1.01

\begin{thebibliography}{30}
% BibTex style file: bmc-mathphys.bst (version 2.1), 2014-07-24
\ifx \bisbn   \undefined \def \bisbn  #1{ISBN #1}\fi
\ifx \binits  \undefined \def \binits#1{#1}\fi
\ifx \bauthor  \undefined \def \bauthor#1{#1}\fi
\ifx \batitle  \undefined \def \batitle#1{#1}\fi
\ifx \bjtitle  \undefined \def \bjtitle#1{#1}\fi
\ifx \bvolume  \undefined \def \bvolume#1{\textbf{#1}}\fi
\ifx \byear  \undefined \def \byear#1{#1}\fi
\ifx \bissue  \undefined \def \bissue#1{#1}\fi
\ifx \bfpage  \undefined \def \bfpage#1{#1}\fi
\ifx \blpage  \undefined \def \blpage #1{#1}\fi
\ifx \burl  \undefined \def \burl#1{\textsf{#1}}\fi
\ifx \doiurl  \undefined \def \doiurl#1{\url{https://doi.org/#1}}\fi
\ifx \betal  \undefined \def \betal{\textit{et al.}}\fi
\ifx \binstitute  \undefined \def \binstitute#1{#1}\fi
\ifx \binstitutionaled  \undefined \def \binstitutionaled#1{#1}\fi
\ifx \bctitle  \undefined \def \bctitle#1{#1}\fi
\ifx \beditor  \undefined \def \beditor#1{#1}\fi
\ifx \bpublisher  \undefined \def \bpublisher#1{#1}\fi
\ifx \bbtitle  \undefined \def \bbtitle#1{#1}\fi
\ifx \bedition  \undefined \def \bedition#1{#1}\fi
\ifx \bseriesno  \undefined \def \bseriesno#1{#1}\fi
\ifx \blocation  \undefined \def \blocation#1{#1}\fi
\ifx \bsertitle  \undefined \def \bsertitle#1{#1}\fi
\ifx \bsnm \undefined \def \bsnm#1{#1}\fi
\ifx \bsuffix \undefined \def \bsuffix#1{#1}\fi
\ifx \bparticle \undefined \def \bparticle#1{#1}\fi
\ifx \barticle \undefined \def \barticle#1{#1}\fi
\bibcommenthead
\ifx \bconfdate \undefined \def \bconfdate #1{#1}\fi
\ifx \botherref \undefined \def \botherref #1{#1}\fi
\ifx \url \undefined \def \url#1{\textsf{#1}}\fi
\ifx \bchapter \undefined \def \bchapter#1{#1}\fi
\ifx \bbook \undefined \def \bbook#1{#1}\fi
\ifx \bcomment \undefined \def \bcomment#1{#1}\fi
\ifx \oauthor \undefined \def \oauthor#1{#1}\fi
\ifx \citeauthoryear \undefined \def \citeauthoryear#1{#1}\fi
\ifx \endbibitem  \undefined \def \endbibitem {}\fi
\ifx \bconflocation  \undefined \def \bconflocation#1{#1}\fi
\ifx \arxivurl  \undefined \def \arxivurl#1{\textsf{#1}}\fi
\csname PreBibitemsHook\endcsname

%%% 1
\bibitem[\protect\citeauthoryear{Atat{\"u}re et~al.}{2018}]{atature2018material}
\begin{barticle}
\bauthor{\bsnm{Atat{\"u}re}, \binits{M.}},
\bauthor{\bsnm{Englund}, \binits{D.}},
\bauthor{\bsnm{Vamivakas}, \binits{N.}},
\bauthor{\bsnm{Lee}, \binits{S.-Y.}},
\bauthor{\bsnm{Wrachtrup}, \binits{J.}}:
\batitle{{Material platforms for spin-based photonic quantum technologies}}.
\bjtitle{Nature Reviews Materials}
\bvolume{3}(\bissue{5}),
\bfpage{38}--\blpage{51}
(\byear{2018})
\end{barticle}
\endbibitem

%%% 2
\bibitem[\protect\citeauthoryear{Shandilya et~al.}{2022}]{shandilya2022diamond}
\begin{barticle}
\bauthor{\bsnm{Shandilya}, \binits{P.K.}},
\bauthor{\bsnm{Fl{\aa}gan}, \binits{S.}},
\bauthor{\bsnm{Carvalho}, \binits{N.C.}},
\bauthor{\bsnm{Zohari}, \binits{E.}},
\bauthor{\bsnm{Kavatamane}, \binits{V.K.}},
\bauthor{\bsnm{Losby}, \binits{J.E.}},
\bauthor{\bsnm{Barclay}, \binits{P.E.}}:
\batitle{{Diamond integrated quantum nanophotonics: spins, photons and phonons}}.
\bjtitle{Journal of Lightwave Technology}
\bvolume{40}(\bissue{23}),
\bfpage{7538}--\blpage{7571}
(\byear{2022})
\end{barticle}
\endbibitem

%%% 3
\bibitem[\protect\citeauthoryear{Orphal-Kobin et~al.}{2025}]{orphal2024coherent}
\begin{barticle}
\bauthor{\bsnm{Orphal-Kobin}, \binits{L.}},
\bauthor{\bsnm{Torun}, \binits{C.G.}},
\bauthor{\bsnm{Bopp}, \binits{J.M.}},
\bauthor{\bsnm{Pieplow}, \binits{G.}},
\bauthor{\bsnm{Schröder}, \binits{T.}}:
\batitle{{Coherent Microwave, Optical, and Mechanical Quantum Control of Spin Qubits in Diamond}}.
\bjtitle{Advanced Quantum Technologies}
\bvolume{8}(\bissue{2}),
\bfpage{2300432}
(\byear{2025})
\end{barticle}
\endbibitem

%%% 4
\bibitem[\protect\citeauthoryear{Hausmann et~al.}{2010}]{hausmann2010fabrication}
\begin{barticle}
\bauthor{\bsnm{Hausmann}, \binits{B.J.}},
\bauthor{\bsnm{Khan}, \binits{M.}},
\bauthor{\bsnm{Zhang}, \binits{Y.}},
\bauthor{\bsnm{Babinec}, \binits{T.M.}},
\bauthor{\bsnm{Martinick}, \binits{K.}},
\bauthor{\bsnm{McCutcheon}, \binits{M.}},
\bauthor{\bsnm{Hemmer}, \binits{P.R.}},
\bauthor{\bsnm{Lon{\v{c}}ar}, \binits{M.}}:
\batitle{{Fabrication of diamond nanowires for quantum information processing applications}}.
\bjtitle{Diamond and Related Materials}
\bvolume{19}(\bissue{5-6}),
\bfpage{621}--\blpage{629}
(\byear{2010})
\end{barticle}
\endbibitem

%%% 5
\bibitem[\protect\citeauthoryear{Babinec et~al.}{2010}]{babinec2010diamond}
\begin{barticle}
\bauthor{\bsnm{Babinec}, \binits{T.M.}},
\bauthor{\bsnm{Hausmann}, \binits{B.J.}},
\bauthor{\bsnm{Khan}, \binits{M.}},
\bauthor{\bsnm{Zhang}, \binits{Y.}},
\bauthor{\bsnm{Maze}, \binits{J.R.}},
\bauthor{\bsnm{Hemmer}, \binits{P.R.}},
\bauthor{\bsnm{Lon{\v{c}}ar}, \binits{M.}}:
\batitle{{A diamond nanowire single-photon source}}.
\bjtitle{Nature nanotechnology}
\bvolume{5}(\bissue{3}),
\bfpage{195}--\blpage{199}
(\byear{2010})
\end{barticle}
\endbibitem

%%% 6
\bibitem[\protect\citeauthoryear{Appel et~al.}{2016}]{appel2016fabrication}
\begin{botherref}
\oauthor{\bsnm{Appel}, \binits{P.}},
\oauthor{\bsnm{Neu}, \binits{E.}},
\oauthor{\bsnm{Ganzhorn}, \binits{M.}},
\oauthor{\bsnm{Barfuss}, \binits{A.}},
\oauthor{\bsnm{Batzer}, \binits{M.}},
\oauthor{\bsnm{Gratz}, \binits{M.}},
\oauthor{\bsnm{Tsch{\"o}pe}, \binits{A.}},
\oauthor{\bsnm{Maletinsky}, \binits{P.}}:
{Fabrication of all diamond scanning probes for nanoscale magnetometry}.
Review of Scientific Instruments
\textbf{87}(6)
(2016)
\end{botherref}
\endbibitem

%%% 7
\bibitem[\protect\citeauthoryear{Neu et~al.}{2014}]{neu2014photonic}
\begin{botherref}
\oauthor{\bsnm{Neu}, \binits{E.}},
\oauthor{\bsnm{Appel}, \binits{P.}},
\oauthor{\bsnm{Ganzhorn}, \binits{M.}},
\oauthor{\bsnm{Miguel-S{\'a}nchez}, \binits{J.}},
\oauthor{\bsnm{Lesik}, \binits{M.}},
\oauthor{\bsnm{Mille}, \binits{V.}},
\oauthor{\bsnm{Jacques}, \binits{V.}},
\oauthor{\bsnm{Tallaire}, \binits{A.}},
\oauthor{\bsnm{Achard}, \binits{J.}},
\oauthor{\bsnm{Maletinsky}, \binits{P.}}:
{Photonic nano-structures on (111)-oriented diamond}.
Applied Physics Letters
\textbf{104}(15)
(2014)
\end{botherref}
\endbibitem

%%% 8
\bibitem[\protect\citeauthoryear{Qnami and QZabre}{2021,2023}]{Qnami2021}
\begin{botherref}
\oauthor{\bsnm{Qnami}},
\oauthor{\bsnm{QZabre}}:
{Quantilever by Qnami and Quantum Scanning Tips by QZabre}.
Borchure.
\url{https://qnami.ch/portfolio/quantilever-mx/} and \url{https://qzabre.com/en/products/scanning-tips}
(2021,2023)
\end{botherref}
\endbibitem

%%% 9
\bibitem[\protect\citeauthoryear{Hessel et~al.}{2006}]{hessel2006hydrogen}
\begin{barticle}
\bauthor{\bsnm{Hessel}, \binits{C.M.}},
\bauthor{\bsnm{Henderson}, \binits{E.J.}},
\bauthor{\bsnm{Veinot}, \binits{J.G.}}:
\batitle{{Hydrogen silsesquioxane: a molecular precursor for nanocrystalline Si-SiO2 composites and freestanding hydride-surface-terminated silicon nanoparticles}}.
\bjtitle{Chemistry of materials}
\bvolume{18}(\bissue{26}),
\bfpage{6139}--\blpage{6146}
(\byear{2006})
\end{barticle}
\endbibitem

%%% 10
\bibitem[\protect\citeauthoryear{Barry et~al.}{2011}]{barry2011synthesis}
\begin{barticle}
\bauthor{\bsnm{Barry}, \binits{S.D.}},
\bauthor{\bsnm{Yang}, \binits{Z.}},
\bauthor{\bsnm{Kelly}, \binits{J.A.}},
\bauthor{\bsnm{Henderson}, \binits{E.J.}},
\bauthor{\bsnm{Veinot}, \binits{J.G.}}:
\batitle{{Synthesis of Si--x Ge1--x Nanocrystals Using Hydrogen Silsesquioxane and Soluble Germanium Diiodide Complexes}}.
\bjtitle{Chemistry of Materials}
\bvolume{23}(\bissue{22}),
\bfpage{5096}--\blpage{5103}
(\byear{2011})
\end{barticle}
\endbibitem

%%% 11
\bibitem[\protect\citeauthoryear{Milliken et~al.}{2021}]{milliken2021tailoring}
\begin{barticle}
\bauthor{\bsnm{Milliken}, \binits{S.}},
\bauthor{\bsnm{Cui}, \binits{K.}},
\bauthor{\bsnm{Klein}, \binits{B.A.}},
\bauthor{\bsnm{Cheong}, \binits{I.T.}},
\bauthor{\bsnm{Yu}, \binits{H.}},
\bauthor{\bsnm{Michaelis}, \binits{V.K.}},
\bauthor{\bsnm{Veinot}, \binits{J.G.}}:
\batitle{{Tailoring B-doped silicon nanocrystal surface chemistry via phosphorus pentachloride--mediated surface alkoxylation}}.
\bjtitle{Nanoscale}
\bvolume{13}(\bissue{43}),
\bfpage{18281}--\blpage{18292}
(\byear{2021})
\end{barticle}
\endbibitem

%%% 12
\bibitem[\protect\citeauthoryear{Mollard et~al.}{2002}]{mollard2002hsq}
\begin{barticle}
\bauthor{\bsnm{Mollard}, \binits{L.}},
\bauthor{\bsnm{Cunge}, \binits{G.}},
\bauthor{\bsnm{Tedesco}, \binits{S.}},
\bauthor{\bsnm{Dal’zotto}, \binits{B.}},
\bauthor{\bsnm{Foucher}, \binits{J.}}:
\batitle{{HSQ hybrid lithography for 20 nm CMOS devices development}}.
\bjtitle{Microelectronic engineering}
\bvolume{61},
\bfpage{755}--\blpage{761}
(\byear{2002})
\end{barticle}
\endbibitem

%%% 13
\bibitem[\protect\citeauthoryear{van Delft et~al.}{2000}]{van2000hydrogen}
\begin{barticle}
\bauthor{\bsnm{Delft}, \binits{F.C.}},
\bauthor{\bsnm{Weterings}, \binits{J.P.}},
\bauthor{\bsnm{Langen-Suurling}, \binits{A.K.}},
\bauthor{\bsnm{Romijn}, \binits{H.}}:
\batitle{{Hydrogen silsesquioxane/novolak bilayer resist for high aspect ratio nanoscale electron-beam lithography}}.
\bjtitle{Journal of Vacuum Science \& Technology B: Microelectronics and Nanometer Structures Processing, Measurement, and Phenomena}
\bvolume{18}(\bissue{6}),
\bfpage{3419}--\blpage{3423}
(\byear{2000})
\end{barticle}
\endbibitem

%%% 14
\bibitem[\protect\citeauthoryear{Trellenkamp et~al.}{2003}]{trellenkamp2003patterning}
\begin{barticle}
\bauthor{\bsnm{Trellenkamp}, \binits{S.}},
\bauthor{\bsnm{Moers}, \binits{J.}},
\bauthor{\bsnm{{van der Hart}}, \binits{A.}},
\bauthor{\bsnm{Kordoš}, \binits{P.}},
\bauthor{\bsnm{Lüth}, \binits{H.}}:
\batitle{{Patterning of 25-nm-wide silicon webs with an aspect ratio of 13}}.
\bjtitle{Microelectronic Engineering}
\bvolume{67-68},
\bfpage{376}--\blpage{380}
(\byear{2003})
\end{barticle}
\endbibitem

%%% 15
\bibitem[\protect\citeauthoryear{Lauvernier et~al.}{2005}]{lauvernier2005realization}
\begin{barticle}
\bauthor{\bsnm{Lauvernier}, \binits{D.}},
\bauthor{\bsnm{Garidel}, \binits{S.}},
\bauthor{\bsnm{Legrand}, \binits{C.}},
\bauthor{\bsnm{Vilcot}, \binits{J.-P.}}:
\batitle{{Realization of sub-micron patterns on GaAs using a {HSQ} etching mask}}.
\bjtitle{Microelectronic engineering}
\bvolume{77}(\bissue{3-4}),
\bfpage{210}--\blpage{216}
(\byear{2005})
\end{barticle}
\endbibitem

%%% 16
\bibitem[\protect\citeauthoryear{Dylewicz et~al.}{2010}]{dylewicz2010fabrication}
\begin{barticle}
\bauthor{\bsnm{Dylewicz}, \binits{R.}},
\bauthor{\bsnm{De~La~Rue}, \binits{R.}},
\bauthor{\bsnm{Wasielewski}, \binits{R.}},
\bauthor{\bsnm{Mazur}, \binits{P.}},
\bauthor{\bsnm{Mez{\H{o}}si}, \binits{G.}},
\bauthor{\bsnm{Bryce}, \binits{A.}}:
\batitle{{Fabrication of submicron-sized features in InP/InGaAsP/AlGaInAs quantum well heterostructures by optimized inductively coupled plasma etching with Cl2/Ar/N2 chemistry}}.
\bjtitle{Journal of Vacuum Science \& Technology B}
\bvolume{28}(\bissue{4}),
\bfpage{882}--\blpage{890}
(\byear{2010})
\end{barticle}
\endbibitem

%%% 17
\bibitem[\protect\citeauthoryear{Gnan et~al.}{2008}]{gnan2008fabrication}
\begin{barticle}
\bauthor{\bsnm{Gnan}, \binits{M.}},
\bauthor{\bsnm{Thoms}, \binits{S.}},
\bauthor{\bsnm{Macintyre}, \binits{D.}},
\bauthor{\bsnm{De~La~Rue}, \binits{R.}},
\bauthor{\bsnm{Sorel}, \binits{M.}}:
\batitle{{Fabrication of low-loss photonic wires in silicon-on-insulator using hydrogen silsesquioxane electron-beam resist}}.
\bjtitle{Electronics Letters}
\bvolume{44}(\bissue{2}),
\bfpage{115}--\blpage{116}
(\byear{2008})
\end{barticle}
\endbibitem

%%% 18
\bibitem[\protect\citeauthoryear{Solard et~al.}{2020}]{solard2020optimal}
\begin{botherref}
\oauthor{\bsnm{Solard}, \binits{J.}},
\oauthor{\bsnm{Chakaroun}, \binits{M.}},
\oauthor{\bsnm{Boudrioua}, \binits{A.}}:
{Optimal design and fabrication of ITO photonic crystal using e-beam patterned hydrogen silsesquioxane resist}.
Journal of Vacuum Science \& Technology B
\textbf{38}(2)
(2020)
\end{botherref}
\endbibitem

%%% 19
\bibitem[\protect\citeauthoryear{Nam et~al.}{2007}]{nam2007electron}
\begin{barticle}
\bauthor{\bsnm{Nam}, \binits{S.-W.}},
\bauthor{\bsnm{Lee}, \binits{T.-Y.}},
\bauthor{\bsnm{Wi}, \binits{J.-S.}},
\bauthor{\bsnm{Lee}, \binits{D.}},
\bauthor{\bsnm{Lee}, \binits{H.-S.}},
\bauthor{\bsnm{Jin}, \binits{K.-B.}},
\bauthor{\bsnm{Lee}, \binits{M.-H.}},
\bauthor{\bsnm{Kim}, \binits{H.-M.}},
\bauthor{\bsnm{Kim}, \binits{K.-B.}}:
\batitle{{Electron-beam lithography patterning of Ge2Sb2Te5 nanostructures using hydrogen silsesquioxane and amorphous Si intermediate layer}}.
\bjtitle{Journal of the Electrochemical Society}
\bvolume{154}(\bissue{9}),
\bfpage{844}
(\byear{2007})
\end{barticle}
\endbibitem

%%% 20
\bibitem[\protect\citeauthoryear{Radtke et~al.}{2019a}]{radtke2019plasma}
\begin{barticle}
\bauthor{\bsnm{Radtke}, \binits{M.}},
\bauthor{\bsnm{Render}, \binits{L.}},
\bauthor{\bsnm{Nelz}, \binits{R.}},
\bauthor{\bsnm{Neu}, \binits{E.}}:
\batitle{{Plasma treatments and photonic nanostructures for shallow nitrogen vacancy centers in diamond}}.
\bjtitle{Optical Materials Express}
\bvolume{9}(\bissue{12}),
\bfpage{4716}--\blpage{4733}
(\byear{2019})
\end{barticle}
\endbibitem

%%% 21
\bibitem[\protect\citeauthoryear{Radtke et~al.}{2019b}]{radtke2019reliable}
\begin{barticle}
\bauthor{\bsnm{Radtke}, \binits{M.}},
\bauthor{\bsnm{Nelz}, \binits{R.}},
\bauthor{\bsnm{Slablab}, \binits{A.}},
\bauthor{\bsnm{Neu}, \binits{E.}}:
\batitle{{Reliable nanofabrication of single-crystal diamond photonic nanostructures for nanoscale sensing}}.
\bjtitle{Micromachines}
\bvolume{10}(\bissue{11}),
\bfpage{718}
(\byear{2019})
\end{barticle}
\endbibitem

%%% 22
\bibitem[\protect\citeauthoryear{Chu et~al.}{2014}]{chu2014coherent}
\begin{barticle}
\bauthor{\bsnm{Chu}, \binits{Y.}},
\bauthor{\bsnm{Leon}, \binits{N.P.}},
\bauthor{\bsnm{Shields}, \binits{B.J.}},
\bauthor{\bsnm{Hausmann}, \binits{B.}},
\bauthor{\bsnm{Evans}, \binits{R.}},
\bauthor{\bsnm{Togan}, \binits{E.}},
\bauthor{\bsnm{Burek}, \binits{M.J.}},
\bauthor{\bsnm{Markham}, \binits{M.}},
\bauthor{\bsnm{Stacey}, \binits{A.}},
\bauthor{\bsnm{Zibrov}, \binits{A.S.}},
\bauthor{\bsnm{Yacoby}, \binits{A.}},
\bauthor{\bsnm{Twitchen}, \binits{D.J.}},
\bauthor{\bsnm{Loncar}, \binits{M.}},
\bauthor{\bsnm{Park}, \binits{H.}},
\bauthor{\bsnm{Maletinsky}, \binits{P.}},
\bauthor{\bsnm{Lukin}, \binits{M.D.}}:
\batitle{{Coherent Optical Transitions in Implanted Nitrogen Vacancy Centers}}.
\bjtitle{Nano Letters}
\bvolume{14}(\bissue{4}),
\bfpage{1982}--\blpage{1986}
(\byear{2014})
\end{barticle}
\endbibitem

%%% 23
\bibitem[\protect\citeauthoryear{Rani et~al.}{2020}]{rani2020recent}
\begin{barticle}
\bauthor{\bsnm{Rani}, \binits{D.}},
\bauthor{\bsnm{Opaluch}, \binits{O.R.}},
\bauthor{\bsnm{Neu}, \binits{E.}}:
\batitle{{Recent advances in single crystal diamond device fabrication for photonics, sensing and nanomechanics}}.
\bjtitle{Micromachines}
\bvolume{12}(\bissue{1}),
\bfpage{36}
(\byear{2020})
\end{barticle}
\endbibitem

%%% 24
\bibitem[\protect\citeauthoryear{Campbell}{2013}]{campbell2008fabrication}
\begin{bbook}
\bauthor{\bsnm{Campbell}, \binits{S.A.}}:
\bbtitle{{Fabrication Engineering at the Micro- and Nanoscale}},
\bedition{4th ed} edn.
\bsertitle{The Oxford series in electrical and computer engineering}.
\bpublisher{Oxford University Press}, \blocation{???}
(\byear{2013})
\end{bbook}
\endbibitem

%%% 25
\bibitem[\protect\citeauthoryear{Frye and Collins}{1970}]{frye1970oligomeric}
\begin{barticle}
\bauthor{\bsnm{Frye}, \binits{C.L.}},
\bauthor{\bsnm{Collins}, \binits{W.T.}}:
\batitle{{Oligomeric silsesquioxanes,(HSiO3/2)n}}.
\bjtitle{Journal of the American Chemical Society}
\bvolume{92}(\bissue{19}),
\bfpage{5586}--\blpage{5588}
(\byear{1970})
\end{barticle}
\endbibitem

%%% 26
\bibitem[\protect\citeauthoryear{Pandey}{2022}]{no2022}
\begin{botherref}
\oauthor{\bsnm{Pandey}, \binits{N.O.}}:
{Quantum Optimal Control for Quantum Sensing with Nitrogen-vacancy Centers}.
PhD thesis,
Technische Universit{\"a}t Kaiserslautern
(2022)
\end{botherref}
\endbibitem

%%% 27
\bibitem[\protect\citeauthoryear{Opaluch et~al.}{2021}]{opaluch2021optimized}
\begin{barticle}
\bauthor{\bsnm{Opaluch}, \binits{O.R.}},
\bauthor{\bsnm{Oshnik}, \binits{N.}},
\bauthor{\bsnm{Nelz}, \binits{R.}},
\bauthor{\bsnm{Neu}, \binits{E.}}:
\batitle{{Optimized planar microwave antenna for nitrogen vacancy center based sensing applications}}.
\bjtitle{Nanomaterials}
\bvolume{11}(\bissue{8}),
\bfpage{2108}
(\byear{2021})
\end{barticle}
\endbibitem

%%% 28
\bibitem[\protect\citeauthoryear{Xiao et~al.}{2015}]{xiao2015fluorescence}
\begin{barticle}
\bauthor{\bsnm{Xiao}, \binits{J.}},
\bauthor{\bsnm{Liu}, \binits{P.}},
\bauthor{\bsnm{Li}, \binits{L.}},
\bauthor{\bsnm{Yang}, \binits{G.}}:
\batitle{{Fluorescence origin of nanodiamonds}}.
\bjtitle{The Journal of Physical Chemistry C}
\bvolume{119}(\bissue{4}),
\bfpage{2239}--\blpage{2248}
(\byear{2015})
\end{barticle}
\endbibitem

%%% 29
\bibitem[\protect\citeauthoryear{Padrez et~al.}{2023}]{padrez2023nanodiamond}
\begin{barticle}
\bauthor{\bsnm{Padrez}, \binits{Y.}},
\bauthor{\bsnm{Golubewa}, \binits{L.}},
\bauthor{\bsnm{Bahdanava}, \binits{A.}},
\bauthor{\bsnm{Jankunec}, \binits{M.}},
\bauthor{\bsnm{Matulaitiene}, \binits{I.}},
\bauthor{\bsnm{Semenov}, \binits{D.}},
\bauthor{\bsnm{Karpicz}, \binits{R.}},
\bauthor{\bsnm{Kulahava}, \binits{T.}},
\bauthor{\bsnm{Svirko}, \binits{Y.}},
\bauthor{\bsnm{Kuzhir}, \binits{P.}}:
\batitle{{Nanodiamond surface as a photoluminescent pH sensor}}.
\bjtitle{Nanotechnology}
\bvolume{34}(\bissue{19}),
\bfpage{195702}
(\byear{2023})
\end{barticle}
\endbibitem

%%% 30
\bibitem[\protect\citeauthoryear{Nelz et~al.}{2019}]{nelz2019toward}
\begin{barticle}
\bauthor{\bsnm{Nelz}, \binits{R.}},
\bauthor{\bsnm{Görlitz}, \binits{J.}},
\bauthor{\bsnm{Herrmann}, \binits{D.}},
\bauthor{\bsnm{Slablab}, \binits{A.}},
\bauthor{\bsnm{Challier}, \binits{M.}},
\bauthor{\bsnm{Radtke}, \binits{M.}},
\bauthor{\bsnm{Fischer}, \binits{M.}},
\bauthor{\bsnm{Gsell}, \binits{S.}},
\bauthor{\bsnm{Schreck}, \binits{M.}},
\bauthor{\bsnm{Becher}, \binits{C.}},
\bauthor{\bsnm{Neu}, \binits{E.}}:
\batitle{{Toward wafer-scale diamond nano- and quantum technologies}}.
\bjtitle{APL Materials}
\bvolume{7}(\bissue{1}),
\bfpage{011108}
(\byear{2019})
\end{barticle}
\endbibitem

\end{thebibliography}

\end{document}